\documentstyle[prd,aps,epsfig]{revtex}

\newcommand{\be}{\begin{equation}}
\newcommand{\ee}{\end{equation}}
\newcommand{\ba}{\begin{eqnarray}}
\newcommand{\ea}{\end{eqnarray}}

\begin{document}

\title{STRONGLY AND WEAKLY UNSTABLE \\[2mm] ANISOTROPIC QUARK-GLUON PLASMA}

\author{Cristina Manuel\footnote{Electronic address:
{\tt cristina.manuel@ific.uv.es}}}

\address{\it Instituto de F\'{\i}sica Corpuscular \\
C.S.I.C.-Universitat de Val\`encia\\
Edificio de Institutos de Paterna, Apt 2085 \\
46071 Val\`encia, Spain}

\author{Stanis\l aw Mr\' owczy\' nski\footnote{Electronic address:
{\tt mrow@fuw.edu.pl}}}

\address{\it So\l tan Institute for Nuclear Studies \\
ul. Ho\.za 69, PL - 00-681 Warsaw, Poland \\
and Institute of Physics, \'Swi\c etokrzyska Academy \\
ul. \'Swi\c etokrzyska 15, PL - 25-406 Kielce, Poland}

\date{6-th July 2005}

\maketitle

\begin{abstract}

Using explicit solutions of the QCD transport equations, we derive 
an effective potential for an anisotropic quark-gluon plasma which
under plausible assumptions holds beyond the Hard Loop approximation.
The configurations, which are unstable in the linear response approach, 
are characterized by a negative quadratic term of the effective 
potential. The signs of higher order terms can be either negative or 
positive, depending on the parton momentum distribution. In the case 
of a Gaussian momentum distribution, the potential is negative and 
unbound from below. Therefore, the modes, which are unstable for 
gauge fields of small amplitude, remain unstable for arbitrary large 
amplitudes. We also present an example of a momentum distribution 
which gives a negative quadratic term of the effective potential 
but the whole potential has a minimum and it grows for sufficiently 
large gauge fields. Then, the system is weakly unstable. The character 
of the instability is important for the dynamical evolution of the 
plasma system.

\end{abstract}




\section{Introduction}


An anisotropic quark-gluon plasma is unstable with respect to the 
transverse mode known as the filamentation or Weibel instability 
\cite{Mrowczynski:xv,Mrowczynski:1996vh,Randrup:2003cw,Romatschke:2003ms,Romatschke:2004jh,Arnold:2003rq}. The instabilities have recently focused some
attention in the context of the phenomenology of nucleus-nucleus collisions
at high energies. The experimental data on the particle spectra and the 
so-called elliptic flow, which have been obtained at the Relativistic 
Heavy-Ion Collider (RHIC) in Brookhaven National Laboratory, suggest, when 
analyzed within the hydrodynamical model, that an equilibration time of the 
parton\footnote{The term `parton' is used to denote a fermionic (quark) or 
bosonic (gluon) excitation of the quark-gluon plasma.} system produced at 
the collision early stage is as short as 0.6 ${\rm fm}/c$ 
\cite{Heinz:2004pj}. Calculations, which assume that the parton-parton 
collisions are responsible for the equilibration of the weakly interacting 
plasma, provide a significantly longer time of at least 2.6 ${\rm fm}/c$ 
\cite{Baier:2002bt}. To thermalize the system one needs either a few hard 
collisions of the momentum transfer of order of the characteristic parton 
momentum\footnote{Although we consider anisotropic systems, the 
characteristic momentum in all directions is assumed to be of the same 
order.}, which we denote here as $T$ (as the temperature of equilibrium 
system), or many collisions of smaller transfer. As discussed in 
{\it e.g.} \cite{Arnold:1998cy}, the time scale of the collisional 
equilibration is of order 
\be
\label{hard-time}
t_{\rm hard} \sim \frac{1}{g^4 {\rm ln}(1/g)\,T} \;, 
\ee
where $g$ is the QCD coupling constant. The characteristic time 
of instability growth is roughly of order $1/gT$ for a sufficiently 
anisotropic momentum distribution
\cite{Mrowczynski:xv,Romatschke:2003ms,Arnold:2003rq,Arnold:2004ti}.
Therefore, the instabilities are much `faster' than the hard collisions
in the weak coupling regime. Since the system's momentum distribution
becomes more isotropic when the unstable modes develop, the instabilities 
have been argued \cite{Mrowczynski:xv,Arnold:2004ti} to effectively speed 
up the thermalization processes. Very recent numerical simulation
\cite{Dumitru:2005gp}, where the fields, particles and color charges 
are treated classically, fully confirms the argument.

We note that the isotropization should be distinguished from the 
equilibration process\cite{Mrowczynski:xv}. The instabilities driven 
isotropization is a mean-field reversible phenomenon which is not 
associated with  entropy production. Therefore, the collisions,
which are responsible for the dissipation, are needed to reach the 
equilibrium state of maximal entropy. The instabilities contribute 
to the equilibration indirectly, reducing relative parton momenta 
and increasing the collision rate. And recently, it has been observed 
that the hydrodynamic collective behavior, which is evident in the
experimental data \cite{Heinz:2004pj}, does not actually require 
local thermodynamic equilibrium but a merely isotropic momentum 
distribution of liquid components \cite{Arnold:2004ti}.

The stability analysis of the plasma system is usually performed 
within a linear response approach which assumes smallness of the 
gauge field amplitudes. An elegant qualitative argument 
\cite{Arnold:2004ih} suggests that non-Abelian non-linearities do 
not stabilize the unstable modes as the system spontaneously chooses 
an Abelian configuration in the course of the instability development. 
This happens because the Abelian configuration corresponds to the 
steepest decrease of the effective Hard Loop potential. However, 
the phenomenon of abelianization which emerges from the numerical 
simulations \cite{Dumitru:2005gp,Arnold:2004ih,Rebhan:2004ur,Arnold:2005vb,Rebhan:2005re} 
is more complex. In $1+1$ dimensions, {\it i.e.} when the gauge potentials
depend on time and one space variable, the simulations 
\cite{Arnold:2004ih,Rebhan:2004ur} of the Hard Loop anisotropic dynamics
\cite{Mrowczynski:2004kv} and the classical simulation \cite{Dumitru:2005gp}
show that initially the abelianization works well but it is less efficient 
when the fields become non-perturbatively large in the course of 
instability growth. The latter effect is nicely seen in the calculations
\cite{Rebhan:2004ur}. In the simulations \cite{Arnold:2005vb,Rebhan:2005re} 
of the Hard Loop dynamics \cite{Mrowczynski:2004kv} performed in $1+3$ 
dimensions, the abelianization stops working at the non-perturbative scale 
and the growth of field amplitudes changes from exponential to linear.

The question arises what happens beyond the Hard Loop approximation, 
whether the higher-order non-linearites stop the instability growth,
or maybe they make the system even more unstable. In attempt to
address the question, we derive an effective potential of the 
plasma configuration which is known to be unstable at the Hard 
Loop level. We follow the method developed in our earlier paper 
\cite{Manuel:2002pb} which uses explicit solutions of the QCD 
transport equations. To get the solutions, the system under study 
is assumed to be static and translation invariant in two space 
directions. The effective potential, which is obtained, thus 
corresponds to the simulations in $1+1$ dimensions. It holds at 
any order of the gauge coupling but a kind of Abelian approximation,
which is justified due to the effect of abelianization, is needed to 
derive it. The higher order terms of the effective potential, which 
are relevant at sufficiently long times of instability development, 
carry information about the actual character of a plasma configuration. 
Depending of the momentum distribution of partons, the plasma is either 
weakly or strongly unstable. 

Although we do {\em not} even attempt to solve the dynamical problem
of temporal evolution of the unstable plasma -- we merely derive 
the effective potential -- we estimate various time scales of the
instability growth. In particular, we show that the quartic and 
higher order terms of the potential become important at the same
time scale when the fields starts to significantly influence the 
particle's momenta, as explained in the concluding section.

Throughout the article we use the natural units with 
$\hbar = c = k_{\rm B} = 1$. In Section~\ref{method} and in 
Appendix~\ref{appendix1}, where we follow the four-dimensional
notation, we use the metric convention $(1,-1,-1,-1)$ and we distinguish
lower and upper Lorentz indices. In the remaining parts of the paper, 
where we mostly deal with three-vectors, all vectors are contravariant 
and to simplify the notation we do not longer distinguish lower and 
upper indices.


\section{The method}
\label{method}


The distribution functions of quarks $(Q)$, antiquarks $(\bar Q)$, and 
gluons $(G)$ are assumed to satisfy the collisionless transport
equations
\begin{mathletters}
\label{transport}
\ba
p^{\mu} D_{\mu}Q( p,x) + {g \over 2}\: p^{\mu}
\left\{ F_{\mu \nu}(x), {\partial Q( p,x) \over \partial p_{\nu}}\right\} 
&=&  0\;,
\label{transport-q}  \\
p^{\mu} D_{\mu}\bar Q( p,x) - {g \over 2} \: p^{\mu}
\left\{ F_{\mu \nu}(x),
{\partial \bar Q( p,x) \over \partial p_{\nu}}\right\} &=& 0\;,
\label{transport-barq} \\
p^{\mu} {\cal D}_{\mu}G( p,x) + {g \over 2} \: p^{\mu}
\left\{ {\cal F}_{\mu \nu}(x),
{\partial G( p,x) \over \partial p_{\nu}} \right\} &=& 0\;,
\label{transport-gluon}
\ea
\end{mathletters}
$\!\!$where $\{...,...\}$ denotes the anticommutator; the covariant 
derivatives $D_{\mu}$ and ${\cal D}_{\mu}$ act as
$$
D_{\mu} = \partial_{\mu} - ig[A_{\mu}(x),...\; ]\;,\;\;\;\;\;\;\;
{\cal D}_{\mu} = \partial_{\mu} - ig[{\cal A}_{\mu}(x),...\;]\;,
$$
$A_{\mu }$ and ${\cal A}_{\mu }$ being four-potentials
in the fundamental and adjoint representations, respectively,
$$
A^{\mu }(x) = A^{\mu }_a (x) \tau_a \;,\;\;\;\;\;
{\cal A}^{\mu }_{ab}(x) = - if_{abc}A^{\mu }_c (x) \; ,
$$
and $f_{abc}$ are the structure constants of the ${\rm SU}(N_c)$ group.
Since the generators of ${\rm SU}(N_c)$ in the adjoint representation 
are given by $(T_a)_{bc} = - i f_{abc}$, one can also write
${\cal A}^\mu = A^\mu _a T^a$. The field strength tensor in the 
fundamental representation is
$F_{\mu \nu}=\partial_{\mu}A_{\nu} - \partial_{\nu}A_{\mu} -ig
[A_{\mu},A_{\nu}]$, while  ${\cal F}_{\mu \nu}$ denotes the
strength tensor in the adjoint representation.

As already mentioned, the instability of interest is a very fast
phenomenon. The characteristic time of instability development
is not only much shorter than the characteristic time of hard 
parton-parton collisions (\ref{hard-time}) but it is also shorter 
than that of soft collisions which control the dissipation of the 
color degrees of freedom. In equilibrium, the collisions with momentum 
transfer of order $g^2 T$ occur at the time scale \cite{Arnold:1998cy}
\be
\label{soft-time}
t_{\rm soft} \sim \frac{1}{g^2 {\rm ln}(1/g)\,T} \;.
\ee
Since the characteristic time of instability growth, which is 
roughly of order $1/gT$ for sufficiently anisotropic momentum distribution
\cite{Mrowczynski:xv,Romatschke:2003ms,Arnold:2003rq,Arnold:2004ti},
is much shorter than $t_{\rm soft}$, the collision terms of transport 
equations (\ref{transport}) can be neglected.

If a solution of the transport equations is known, one finds the 
associated color current which in the fundamental representation is
\be
\label{current}
j^{\mu }(x) = -\frac{g}{2} \int dP \; p^{\mu}
\Big[ Q( p,x) - \bar Q ( p,x) 
- {1 \over N_c}{\rm Tr}\big[Q(p,x) - \bar Q ( p,x)\big]
+  2i \tau_a f_{abc} G_{bc}( p,x)\Big] \;,
\ee
where the momentum measure
$$
dP = \frac{d^4p}{(2 \pi)^3} 2\Theta(p_0)\, \delta(p^2)
$$
takes into account the mass-shell condition  $p_0 = E_p = |{\bf p}|$.
The adjoint currents equal $j^\mu_a (x) = 2 {\rm Tr} [\tau_a j^\mu (x)]$. 
Since the current is given by the relation
\be
\label{current-action}
j_a^\mu(x) = - \frac{\delta S_{\rm eff}}{\delta A^a_\mu(x)} \;,
\ee
where $S_{\rm eff} \equiv \int d^4x \, {\cal L}_{\rm eff} \;$
is the effective action to be added to the Yang-Mills one, one 
obtains $S_{\rm eff}$ integrating Eq.~(\ref{current-action}).

Finding exact solutions of the transport equations (\ref{transport}) 
is in general a difficult task. However, it is possible to get such 
solutions under some restrictive conditions. Here we consider a system 
where both the gauge field and the distribution functions are 
invariant under some space-time translation(s), {\it i.e.}
\be
\label{condition1}
\partial_{\alpha_i} A^\mu (x) = 0 \ , \qquad  \mu=0,1,2,3   \ ,
\ee
and
\be
\label{homo}
\partial_{\alpha_i}  Q( p,x) =
\partial_{\alpha_i}  \bar Q( p,x) =
\partial_{\alpha_i}  G( p,x) = 0 \;,
\ee
for a fixed $\alpha_i$, which can involve more than one Lorentz 
index. If $\alpha_i = 0$ the system is static while for
$\alpha_i =1,2,3$ the system is homogeneous - the gauge field
and the distribution functions depend only on time.

We look for the solutions of the transport equations 
(\ref{transport}) in the form:
\begin{mathletters}
\label{ansatz}
\ba
\label{ansatz-q}
 Q( p,x) &=& f(p_{\alpha_i} - gA_{\alpha_i}(x) ) =
\sum_{n=0}^\infty \frac{(-g)^n}{n!} A_{\alpha_1}(x)\,
A_{\alpha_2}(x) \cdots A_{\alpha_n}(x)\, \frac{\partial^n
f(p_{\alpha_i})} {\partial p_{\alpha_1}\, \partial p_{\alpha_2}
\ldots \, \partial p_{\alpha_n}} \; ,
\\
\bar Q( p,x) &=& \bar f(p_{\alpha_i} + gA_{\alpha_i}(x) ) =
\sum_{n=0}^\infty \frac{g^n}{n!} A_{\alpha_1}(x)\, A_{\alpha_2}(x)
\cdots A_{\alpha_n}(x)\, \frac{\partial^n \bar f(p_{\alpha_i})}
{\partial p_{\alpha_1}\, \partial p_{\alpha_2} \ldots \partial
p_{\alpha_n}} \;,
\\
G( p,x) &=& f_g(p_{\alpha_i} - g{\cal A}_{\alpha_i}(x) ) =
\sum_{n=0}^\infty \frac{(-g)^n}{n!} {\cal A}_{\alpha_1}(x)\, {\cal
A}_{\alpha_2}(x) \cdots {\cal A}_{\alpha_n}(x)\, \frac{\partial^n
f_g(p_{\alpha_i})} {\partial p_{\alpha_1}\,
\partial p_{\alpha_2} \ldots \partial p_{\alpha_n}} \;,
\label{ansatz-gluon}
\ea
\end{mathletters}
$\!\!$where it is understood that a sum is taken over the
repeated indices. The functions $f$, $\bar f$ and $f_g$ are,
in principle, arbitrary but they can be fixed by additional
considerations. The solutions of the form (\ref{ansatz}) do 
{\em not} assume smallness of $gA$ provided the series 
(\ref{ansatz}) are not terminated at a finite number of 
terms. We stress that all components of the four-vector 
$p^\mu$ are treated here as independent variables.

One finds that the expressions (\ref{ansatz}) exactly solve
Eqs.~(\ref{transport}) when
\be
\label{condition2}
[D_{\mu}A_{\alpha_i}, A_{\alpha_j}] = 0 \ , \qquad
\mu = 0,1,2,3 \; ,
\ee
and
\be
\label{condition2-g}
[{\cal D}_{\mu}{\cal A}_{\alpha_i}, {\cal A}_{\alpha_j}] =0 \, \qquad
\mu = 0,1,2,3 \;.\ee
We note that Eq.~(\ref{condition2-g}) is automatically satisfied
if Eq.~(\ref{condition2}) holds. The condition (\ref{condition2}) 
is satisfied when, in particular, $A^\mu_a$ has only non-vanishing
components in the Cartan subalgebra of ${\rm SU}(N_c)$ which for
${\rm SU}(3)$ are in the $a=3,8$ directions of the color space.
If $A^\mu_a$ has only one color component, the condition 
(\ref{condition2}) is also trivially satisfied.

Having solutions of the transport equations, one finds the associated
color current. Inserting the expressions (\ref{ansatz}) into
Eq.~(\ref{current}), one gets
\ba
\label{current2}
j_a^{\mu}(x) &=& \sum_{n=0}^\infty
\frac{(-g)^{n+1}}{n!} A_{\alpha_1}^{c_1} (x) \cdots
A_{\alpha_n}^{c_n} (x)
 \int dP \;  p^\mu
\\ [1mm] \nonumber
&\times& \bigg\{{\rm Tr} [\tau_a \tau_{c_1} \cdots \tau_{c_n}]
\bigg[\frac{\partial^n f(p_{\alpha_i})} {\partial p_{\alpha_1}
\cdots \,\partial p_{\alpha_n}}
+  (-1)^{n+1} \frac{\partial^n {\bar f}(p_{\alpha_i})} {\partial
p_{\alpha_1} \cdots \,\partial p_{\alpha_n}} \bigg]+ {\rm Tr} [T_a
T_{c_1} \cdots T_{c_n}] \frac{\partial^n  f_g
(p_{\alpha_i})}{\partial p_{\alpha_1} \cdots \,
\partial p_{\alpha_n}} \bigg\}  \;.
\ea The current (\ref{current2}) provides, after integrating
Eq.~(\ref{current-action}), the effective lagrangian of the form 
\cite{Manuel:2002pb}
\ba
\label{efflag}
\nonumber
{\cal L}_{\rm eff} = - \sum_{n=0}^\infty
\frac{(-g)^{n+1}}{(n+1)!}
&\bigg\{& {\rm Tr} [A_{\alpha_1} (x) \cdots
A_{\alpha_{n+1}} (x)]  \int dP \: p^{\alpha_1} \:\bigg[
\frac{\partial^{n}  f(p_{\alpha_i})}{\partial p_{\alpha_2} \cdots\,
\partial p_{\alpha_{n+1}}}
+ (-1)^{n+1} \frac{\partial^n {\bar f}(p_{\alpha_i})}{\partial
p_{\alpha_1} \cdots \,\partial p_{\alpha_n}} \bigg]
\\ [1mm] \label{eff-lagrangian}
&+& {\rm Tr} [{\cal A}_{\alpha_1} (x) \cdots {\cal A}_{\alpha_{n+1}} (x)]
\int dP \: p^{\alpha_1} \:
\frac{\partial^{n}  f_g(p_{\alpha_i})}{\partial p_{\alpha_2} \cdots\,
\partial p_{\alpha_{n+1}}} \bigg\} \;.
\ea
The distribution functions $f$, $\bar f$ and $f_g$ include two helicity
states of, respectively, quarks, antiquarks and gluons. Various quark
flavour states are also included in $f$ and $\bar f$. We note that 
Eq.~(\ref{eff-lagrangian}) with the equilibrium distribution functions 
provides in the static case ($\alpha_i = 0$) the full one-loop 
effective potential \cite{Manuel:2002pb}.

To get the complete action, the lagrangian (\ref{eff-lagrangian})
should be supplemented by the gauge field `kinetic' term
$-\frac{1}{4}F^a_{\mu \nu} F^{a\mu \nu}$ which involves, due to
its non-Abelian nature, not only quadratic but also cubic and 
quartic terms of the potential $A^{\mu}_a$. Since the lagrangian 
(\ref{eff-lagrangian}) has been derived under the assumption that 
the gauge fields obey the condition (\ref{condition2}), the same 
condition must be applied to the `kinetic' term, which then becomes 
effectively Abelian.

As seen, the effective lagrangian (\ref{efflag}) is local. 
In general, this as a consequence of the space-time homogeneity 
(\ref{condition1},\ref{homo}) combined with the commutation condition 
(\ref{condition2}). However, it should be stressed that, as explicitly
shown in \cite{Manuel:2002pb}, the $g^2$ term of the lagrangian 
(\ref{efflag}) does not require the condition (\ref{condition2}) which 
is needed for the higher order terms. The quadratic term and its
relationship to the potential obtained within the linear response
method is further discussed in Appendix~\ref{appendix1}.

The Hard Loop action can be found for both the equilibrium 
\cite{Braaten:1991gm,Frenkel:1991ts,Kelly:1994dh,Blaizot:1993be} 
and anisotropic \cite{Mrowczynski:2004kv} plasmas, modifying the 
$g^2$ term of the lagrangian (\ref{efflag}) to comply with gauge 
invariance. Presumably, the higher order terms of the effective 
potential (\ref{efflag}) can be also generalized in the same way, 
once the space-time invariance is abandoned. However, we are not going 
to follow this direction. Our purpose here is to understand the possible 
role of the higher order terms of Eq.~(\ref{efflag}), whose coefficients 
are uniquely fixed by powers of $g$ and by the parton distribution 
functions.


\section{Effective potential for unstable configurations}
\label{potential}


In this section we discuss configurations which are known to be
unstable in the linear response approach. Specifically, we consider
the filamentation instability which occurs in the homogeneous
plasma when the momentum distribution is anisotropic. When the
instability develops, the kinetic energy of particles is converted
into  field energy. The wave vector of the unstable mode is
perpendicular to the direction with a `surplus' of the momentum
and it points towards the direction with the momentum `deficit'.
We choose the wave vector along the $z$ axis and we expect to
observe an instability when the momentum distribution is elongated
in the $x$ or $y$ direction when compared to the momentum in the $z$ 
direction. In Sec.~\ref{method} we have carefully distinguished the 
upper and lower indices of four-vectors. From now on, all four-vectors 
are contravariant and the upper and lower indices are no longer distinguished
(except the Yang-Mills lagrangian written in a relativistically covariant 
form).

Let us now consider a geometry of the filamentation instability
referring to classical electrodynamics \cite{Mrowczynski:1996vh}.
Due to the `surplus' of momentum along the $x$ or $y$ axis, the
fluctuations of current flowing in the $x\!-\!y$ plane are amplified.
Thus, we expect non-vanishing $x$ or/and $y$ current components.
When the wave vector of a standing (time independent) wave points in
the $z$ direction (${\bf k} = (0,0,k)$) and the current (${\bf j}$)
flows in the $x\!-\!y$ plane, there are non-vanishing $x$ and $y$
components of the vector potential (${\bf A}$) and the magnetic field
(${\bf B}$). Indeed, one checks that
\ba
\nonumber
{\bf j} = (a_x k^2{\rm sin}(kz), a_y k^2{\rm sin}(kz),0) \;, \;\;\;\;\;\;\;
{\bf A} = (a_x{\rm sin}(kz), a_y{\rm sin}(kz),0) \;, \;\;\;\;\;\;\;
{\bf B} = (-a_y k \,{\rm cos}(kz),a_x k \,{\rm cos}(kz),0) \;,
\ea
which depend only on $z$, satisfy the electromagnetic equations
\ba
\nonumber
\nabla \times {\bf B} = {\bf j} \;,\;\;\;\;\;\;\;
\nabla \times {\bf A} = {\bf B} \;.
\ea

As seen, the dynamics of the system, which is static and invariant 
with respect to the translations in $x$ and $y$ directions, can be still
nontrivial. Therefore, we consider such a quark-gluon system. The solution 
of the transport equation is then of the form 
$f(p_0 - gA_0(z), p_x - gA_x(z), p_y - gA_y(z))$. Since the filamentation 
mode is transverse and the magnetic field is responsible for the 
instability development, we can choose the gauge condition $A_0 = 0$. 
Finally, the distribution function is assumed to obey the mirror symmetry
\be
\label{mirror}
f(p_0,{\bf p}) = f(p_0,-{\bf p}) \;.
\ee
Then, the formula (\ref{eff-lagrangian}) for $\alpha_i = x, y$ gives
the effective potential $V_{\rm eff} \equiv - {\cal L}_{\rm eff}$ as
\ba
\label{eff-potential}
V_{\rm eff} & = & - \frac{g^2}{2!} \Big\{
\Big\langle \frac{p_x}{p_0} \frac{\partial}{\partial p_x}
\Big\rangle \, {\rm Tr} [ A_x^2 ] 
+ \Big\langle \frac{p_y}{p_0}
\frac{\partial}{\partial p_y} \Big\rangle \, {\rm Tr} [ A_y^2 ] \Big\}
\nonumber \\ [1mm]
&-& \frac{g^4}{4!} \Big\{ \Big\langle
\frac{p_x}{p_0} \frac{\partial^3}{\partial p^3_x}
\Big\rangle \,{\rm Tr} [ A_x^4] 
+ \Big\langle \frac{p_y}{p_0} \frac{\partial^3}{\partial p^3_y}
\Big\rangle\,{\rm Tr} [ A_y^4]
+ 3 \Big(\Big\langle \frac{p_y}{p_0}
\frac{\partial^3}{\partial p^2_x \partial p_y} \Big\rangle
+ \Big\langle \frac{p_x}{p_0}
\frac{\partial^3}{\partial p^2_y \partial p_x} \Big\rangle
\Big) \,{\rm Tr} [ A_x^2 A_y^2] \Big\} + {\cal O}(g^6) \;,
\ea
where
\be
\langle \cdots \rangle \equiv \int dP \, p^0 \; \cdots f(p) \;.
\ee
If the quantity, which is averaged over four-momenta, is independent 
of $p_0$, the integration over $p_0$ is performed and
$$
\langle \cdots \rangle = \int \frac{d^3p}{(2\pi )^3} \cdots f({\bf p}) \;.
$$
The function $ f({\bf p})$ should be understood as $ f(E_p,{\bf p})$. 
We note that only the quark contribution is taken into account in
Eq.~(\ref{eff-potential}). Including the contributions of antiquarks 
and gluons is straightforward. We also observe that due to the mirror
symmetry condition (\ref{mirror}), there are only terms of even powers 
of $A$ in Eq.~(\ref{eff-potential}).

To study the system's stability, the effective potential generated by the
quasiparticles should be supplemented by a potential-like contribution 
coming from the Yang-Mills `kinetic' contribution to the lagrangian 
$-\frac{1}{4}F^a_{\mu \nu} F^{a \,\mu \nu}$. Since we consider static 
configurations in the $A^0=0$ gauge, the lagrangian equals
$$
- \frac{1}{4}F^a_{\mu \nu} F^{a \,\mu \nu} =
- \frac{1}{2} {\bf B}^a {\bf B}^a \;,
$$
where the chromomagnetic field ${\bf B}^a$ is defined as
$$
{\bf B}^a \equiv \nabla \times {\bf A}^a +
\frac{g}{2} f^{abc} {\bf A}^b \times {\bf A}^c \;.
$$
The Yang-Mills lagrangian is built of the gauge potentials and
their gradients. We identify as a contribution to the effective
potential the term of the Yang-Mills lagrangian where gradients
of ${\bf A}^a$ are absent. This is the only term which survives
in a homogeneous system, and it reads
\be
\label{YM-potential}
V_{\rm YM} = \frac{1}{4} g^2 f^{abc} f^{ade}
({\bf A}^b {\bf A}^d) ({\bf A}^c {\bf A}^e) \;.
\ee
Because of this term, the system has been argued \cite{Arnold:2004ih}
to spontaneously choose an Abelian configuration in the course of the
instability development.  We will assume that this is the case, thus
justifying the Abelian-like conditions (\ref{condition1}) under which 
the effective Lagrangian (\ref{efflag}) was obtained. As noted 
in the Introduction, the simulations in $1+1$ dimensions
\cite{Dumitru:2005gp,Arnold:2004ih,Rebhan:2004ur}, which are relevant 
for our study, confirm effectiveness of the abelianization. We return
to this problem in the concluding section.

The negative sign of the quadratic term of the potential 
(\ref{eff-potential}) usually signals the instability. Let us consider 
the issue. Within the linear response analysis, unstable modes are 
found, see {\it e.g.} \cite{Mrowczynski:2000ed}, as solutions 
of the dispersion equation
$$
{\rm det}[ {\bf k}^2 \delta^{ij} - k^i k^j 
- \omega^2 - \Pi^{ij}(\omega,{\bf k})] = 0 \;.
$$
The polarization tensor of anisotropic plasma, which was derived in 
\cite{Mrowczynski:2000ed}, is given in Appendix~\ref{appendix1} by
Eq.~(\ref{Pi}).

Further, we consider a simplified situation where the wave vector 
points in the $z$ direction (${\bf k} = (0,0,k)$) and the 
chromoelectromagnetic field has only one non-vanishing component 
along the $x$ axis. Then, as discussed in {\it e.g.} \cite{Randrup:2003cw},
the dispersion equation for the mirror symmetric momentum 
distribution simplifies to
\be
\label{dis-eq}
k^2 - \omega^2 - \Pi^{xx}(\omega,k) = 0 \;.
\ee

The unstable modes of interest correspond to the solutions of 
Eq.~(\ref{dis-eq}) with  positive imaginary frequency. The existence 
of unstable modes is controlled by the so-called Penrose criterion 
\cite{Kra73}, which was applied to the simplified equation 
(\ref{dis-eq}) in \cite{Mrowczynski:xv} while a general 
configuration was studied in \cite{Arnold:2003rq}. According 
to the Penrose criterion, there are unstable solutions of 
Eq.~(\ref{dis-eq}) if there are such wave vectors $k$ that
\be
\label{Penrose}
k^2 - \Pi^{xx}(\omega=0,k) < 0 \;.
\ee
Since $\Pi^{xx}(\omega=0,k)$, as given by Eq.~(\ref{Pi}), is 
independent of $k$, the Penrose criterion (\ref{Penrose}) reads
\be
\label{criterion-insta}
\int dP \: p_x
\bigg[ \frac{\partial f}{\partial p_x} 
- \frac{p_x}{p_z} \frac{\partial f}{\partial p_z} 
\bigg] > 0 \;.
\ee
The respective quadratic term of the effective potential 
(\ref{eff-potential}) is negative when
\be
\label{criterion-pot}
\int dP \: p_x
\frac{\partial f}{\partial p_x} > 0 \;.
\ee The criteria (\ref{criterion-insta}, \ref{criterion-pot}) appear
to be fully equivalent to each other. Since the system under 
consideration is static and translationally invariant along the 
$x$ and $y$ directions, the solutions of the transport equations 
cannot explicitly depend on $p_z$. Therefore, the last term in
Eq.~(\ref{criterion-insta}) vanishes identically.

Essential information contained in the potential (\ref{eff-potential}) 
is carried by the sign of each term. As discussed above, the negative 
sign of the quadratic term signals instability of small amplitude modes. 
Depending on signs of higher order terms, one deals with strongly 
unstable or weakly unstable configurations. In the next sections we 
consider examples of momentum distributions which illustrate the two 
situations. 


\section{Strongly unstable configuration}
\label{strong}


We consider here a specific Gaussian form of the distribution
function $f$. Namely, the function is chosen as
\be
\label{gauss}
f(p_x,p_y,p_0) = 2^3 \pi^{3/2}
\sqrt{\beta (\beta - \alpha_x) (\beta -\alpha_y)}
\,\rho \,
{\rm exp}\Big( \alpha_x p_x^2 + \alpha_y p_y^2 - \beta \, p_0^2 \Big) \;,
\ee
where $\rho$ is the quark density ($\rho = \langle 1 \rangle$)
while the parameters $\alpha_x$, $\alpha_y$ and $\beta$ ($\beta > 0 \,, \;
\beta > \alpha_x \,,\; \beta > \alpha_y $) control the momentum
distribution. One observes that taking into account the mass-shell
constraint $p_0 = E_p = \sqrt{p_x^2 + p_y^2 + p_z^2}$, the distribution
function equals
\be
\label{gauss-momentum}
f(p_x,p_y, E_p) =
2^3 \pi^{3/2}
\sqrt{\beta (\beta -\alpha_x) (\beta -\alpha_y)} \,\rho \,
{\rm exp}\Big(-(\beta -\alpha_x)p_x^2 -(\beta -\alpha_y) p^2_y
-\beta  p_z^2 \Big)\;,
\ee
and one immediately finds
$$
\langle p_x^2 \rangle = \frac{\rho}{2(\beta - \alpha_x)} \;, \;\;\;\;\;\;\;
\langle p_y^2 \rangle = \frac{\rho}{2(\beta - \alpha_y)} \;, \;\;\;\;\;\;\;
\langle p_z^2 \rangle = \frac{\rho}{2\beta} \;.
$$

With the distribution function (\ref{gauss}), the momentum
derivatives in Eq.~(\ref{eff-potential}) are trivially computed
and the effective potential reads
\ba
\label{gauss-potential}
V_{\rm eff} &=& - g^2 \Big\{
\alpha_x \Big\langle \frac{p_x^2}{E_p} \Big\rangle \,
{\rm Tr} [A_x^2]
+ \alpha_y \Big\langle \frac{p_y^2}{E_p} \Big\rangle \,
{\rm Tr} [A_y^2] \Big\}
\\ [2mm] \nonumber
&&
- g^4 \bigg\{
\Big(\frac{1}{3} \alpha_x^3  \Big\langle \frac{p_x^4}{E_p} \Big\rangle
+  \frac{1}{2} \alpha_x^2 \Big\langle \frac{p_x^2}{E_p} \Big\rangle \Big)
\,{\rm Tr} [ A_x^4]
+ \Big(\frac{1}{3} \alpha_y^3  \Big\langle \frac{p_y^4}{E_p} \Big\rangle
+  \frac{1}{2} \alpha_y^2 \Big\langle \frac{p_y^2}{E_p} \Big\rangle \Big)
\,{\rm Tr} [ A_y^4] \\ [2mm] \nonumber
&& \;\;\;\;\;\;
+ \Big( (\alpha_x \alpha_y^2 + \alpha_x^2 \alpha_y )
   \Big\langle \frac{p_x^2 p_y^2}{E_p} \Big\rangle
+  \frac{1}{2} \alpha_x \alpha_y
\Big\langle \frac{p_x^2 + p_y^2}{E_p} \Big\rangle
\Big) \,{\rm Tr} [ A_x^2 A_y^2] \bigg\}
+ {\cal O}(g^6) \;.
\ea
The terms of higher powers of $g$, which are suppressed in the small
coupling regime, can be easily calculated as well. The moments of the 
Gaussian distribution function can be computed analytically and expressed 
through  hypergeometric functions but the results are of rather 
complicated form, and thus they are not very useful. We observe that 
the moments of the Gaussian distribution function, which are present 
in Eq.~(\ref{gauss-potential}), are all positive. Therefore, the 
potential (\ref{gauss-potential}) is negative when $\alpha_x$ and 
$\alpha_y$ are positive. We note that the higher order terms are also 
negative for $\alpha_x > 0$ and $\alpha_y > 0$. Then, the effective 
potential is negative and unbound from below for an arbitrary magnitude of 
the gauge potential. We also observe that the infinite series with the 
first few terms given by Eq.~(\ref{gauss-potential}) is convergent for 
any value of the gauge potential as the Gaussian function is regular 
and its derivatives of any order are finite. 

Obviously, the gauge field amplitude cannot roll down to infinity, as 
suggests the potential (\ref{gauss-potential}), and there are several 
dynamical processes that can stop the field growth, rebuilding the 
effective potential. We keep in mind here the non-Abelian non-linearities, 
isotropization, equilibration which operate at different time scales
briefly discussed in the concluding section. Quantitative analysis of 
these effects requires a dynamic treatment which includes the back reaction 
and self-consistent generation of fields. We only mention here that if no 
other processes stop the instability development, the inter-parton 
collisions will certainly do the job. The collisions, which are neglected 
in our analysis, slowly drive the system towards equilibrium. And when 
the thermodynamic equilibrium is reached, the system is no longer 
unstable - the effective potential becomes positive with a minimum 
corresponding the equilibrium configuration. 

Below we consider two special cases of the distribution function
(\ref{gauss}) which are of interest in the context of phenomenology of 
heavy-ion collisions. When two relativistic nuclei of equal mass collide, 
the parton momentum distribution in the center of mass frame is initially 
elongated along the beam axis - it corresponds to a prolate ellipsoid. With 
the time passing, the parton momentum distribution in the local rest 
frame evolves due to the parton free streaming towards an oblate shape with 
the momentum squeezed along the beam direction. The two configurations
are discussed below.

\subsection{Prolate shape}

When $\alpha_y=0$ and $\alpha_\parallel \equiv \alpha_x$, the momentum 
distribution (\ref{gauss-momentum}) is
\be 
\label{gauss-momentum-prolate}
f(p_x,p_y, E_p) = 
2^3 \pi^{3/2} 
\beta \sqrt{\beta -\alpha_\parallel} \,\rho \, {\rm exp}
\Big(-(\beta -\alpha_\parallel)p_x^2 -\beta ( p^2_y + p_z^2 )\Big)\;,
\ee
where, as previously, $\alpha_\parallel < \beta$ and $\beta > 0$, and 
$$ 
\langle p_x^2 \rangle = \frac{\rho}{2(\beta - \alpha_\parallel)} \;, 
\;\;\;\;\;\;\;
\langle p_y^2 \rangle = \langle p_z^2 \rangle = \frac{\rho}{2\beta} \;.
$$
For $\alpha_\parallel > 0$ the distribution (\ref{gauss-momentum-prolate}) 
corresponds to the prolate ellipsoid elongated along the $x$ axis.
The effective potential (\ref{gauss-potential}) for the distribution
(\ref{gauss-momentum-prolate}) equals
\be
\label{gauss-potential-prolate}
V_{\rm eff} = - g^2 
\alpha_\parallel \Big\langle \frac{p_x^2}{E_p} \Big\rangle \,
{\rm Tr} [A_x^2] 
- g^4 \Big(\frac{1}{3} \alpha_\parallel^3  
\Big\langle \frac{p_x^4}{E_p} \Big\rangle
+  \frac{1}{2} \alpha_\parallel^2 
\Big\langle \frac{p_x^2}{E_p} \Big\rangle \Big)
\,{\rm Tr} [A_x^4]  + {\cal O}(g^6) \;.
\ee
One observes that when the momentum ellipsoid is elongated in the $x$
direction ($\alpha_\parallel > 0$), the potential 
(\ref{gauss-potential-prolate}) is negative and unbound from below. 

\subsection{Oblate shape}

When $\alpha_\perp \equiv \alpha_x = \alpha_y$, the momentum 
distribution (\ref{gauss-momentum}) is
\be 
\label{gauss-momentum-oblate}
f(p_x,p_y, E_p) = 
2^3 \pi^{3/2} 
\sqrt{\beta} (\beta -\alpha_\perp) \,\rho \,
{\rm exp}\Big(-(\beta -\alpha_\perp) p_\perp^2  
-\beta  p_z^2 \Big)\;,
\ee
where $\alpha_\perp < \beta$, $\beta > 0$ and $p_\perp^2 \equiv p_x^2 + p_y^2$.
One observes that
$$ 
\langle p_x^2 \rangle = \langle p_y^2 \rangle 
= \frac{1}{2} \langle p_\perp^2 \rangle
= \frac{\rho}{2(\beta - \alpha_\perp)} \;, \;\;\;\;\;\;\;
\langle p_z^2 \rangle = \frac{\rho}{2\beta} \;.
$$
For $\alpha_\perp > 0$ the distribution (\ref{gauss-momentum-oblate}) 
corresponds to the oblate ellipsoid squeezed along the $z$ axis.
The effective potential (\ref{gauss-potential}) for the momentum 
distribution (\ref{gauss-momentum-oblate}) equals
\ba
\label{gauss-potential-oblate}
V_{\rm eff} &=& - \frac{1}{2} g^2
\alpha_\perp \Big\langle \frac{p_\perp^2}{E_p} \Big\rangle \,
{\rm Tr} [A_x^2 + A_y^2] 
\\ [2mm] \nonumber
&& 
- g^4 \bigg\{ 
\Big(\frac{1}{8} \alpha_\perp^3  
\Big\langle \frac{p_\perp^4}{E_p} \Big\rangle
+  \frac{1}{4} \alpha_\perp^2 
\Big\langle \frac{p_\perp^2}{E_p} \Big\rangle \Big)
\,{\rm Tr} [ A_x^4 + A_y^4] 
+ \Big(\frac{1}{4} \alpha_\perp^3 
\Big\langle \frac{p_\perp^4}{E_p} \Big\rangle
+  \frac{1}{2} \alpha_\perp^2  
\Big\langle \frac{p_\perp^2}{E_p} \Big\rangle \Big)
\,{\rm Tr} [ A_x^2 A_y^2] \bigg\}
+ {\cal O}(g^6) \;.
\ea
When the momentum ellipsoid is squeezed in the $z$ direction 
($\alpha_\perp > 0$), the potential (\ref{gauss-potential-oblate})
is negative and unbound from below - all terms are negative.


\section{Weakly unstable configuration}
\label{weak}


The effective potential crucially depends on the parton momentum
distribution. The Gaussian distribution discussed in the previous
section provides an effective potential where the quartic and
higher order terms are negative once the quadratic term is negative. 
In such a case, a mode, which is unstable in the linear response 
approach, remains unstable for gauge fields of larger amplitude. 
However, it is not difficult to invent a momentum distribution 
which gives a negative quadratic term of the effective potential 
and a positive quartic term. Then, an unstable mode in the linear 
response approximation is stabilized when the field's amplitude 
becomes sufficiently large.

As previously, we consider here a static system which is homogeneous
along the $x$ and $y$ directions and we choose the momentum
distribution of the form
\be
\label{stabil-distri}
f(p_x, p_y, p_0) = \frac{2\pi^2 \beta^2}{1 + \beta {\cal P}} \, 
e^{\beta {\cal P}} \, \rho \,
\big[ \delta (p_x - {\cal P}) + \delta (p_x + {\cal P}) \big]
\, e^{- \beta p_0} \;,
\ee
with $\beta > 0$ and ${\cal P} > 0$. As in the case of the Gaussian
distribution, the normalizing constant is chosen in such a way
that the quark density $\rho$ equals $\langle 1 \rangle$. The 
computation of the constant is explained in  
Appendix~\ref{appendix2}. The distribution (\ref{stabil-distri}) 
describes two counter-streaming beams of quarks with the momenta 
$p_x={\cal P}$ and $p_x=- {\cal P}$. The momentum in the $y\!-\!z$ 
plane is distributed according to
$\sim {\rm exp}\Big(-\beta \sqrt{{\cal P}^2 + p_y^2 + p_z^2}\Big)$.

The effective potential (\ref{eff-potential}) computed with the
distribution (\ref{stabil-distri}) equals
\ba
\nonumber
V_{\rm eff} = \frac{g^2}{2!} 
\rho \, \beta \, 
\frac{1 - \beta {\cal P}}{1 + \beta {\cal P}}
\, {\rm Tr} [A_x^2] 
&+& \frac{g^4}{4!} 
\rho \, \beta^3 \, 
\frac{3 - \beta {\cal P}}{1 + \beta {\cal P}}
\, {\rm Tr} [A_x^4] 
\\[2mm] \label{stabil-potential}
&+& \cdots +
\frac{g^{2n}}{(2n)!} 
\rho \, \beta^{2n-1} \, 
\frac{2n - 1 - \beta {\cal P}}{1 + \beta {\cal P}}
\, {\rm Tr} [A_x^{2n}] 
+ \cdots 
\ea
The calculation details are given in Appendix~\ref{appendix2}. 
As seen, for $1/\beta < {\cal P} < 3/\beta$ the quadratic term is 
negative but the quartic and higher term are positive. Then, the mode, 
which is unstable in the linear response approximation, is stabilized 
when the field's amplitude becomes so large that the quartic term matters.

\begin{figure}
\vspace{0.5cm}
\centerline{\epsfig{file=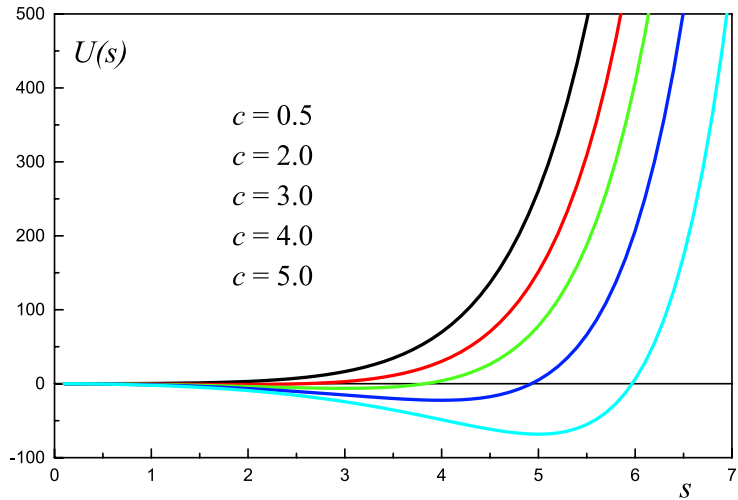,width=100mm}}
\caption{The effective potential (\protect\ref{U-potential})
for 5 values of the parameter $c$. The most upper line
corresponds to $c=0.5$, the lower one to $c=2.0$, etc.} \end{figure}

Interestingly, the series (\ref{stabil-potential}) can be summed up 
and the result is
\be
\label{full-potential}
V_{\rm eff} = \frac{\rho}{\beta (1 + \beta {\cal P})} 
\, {\rm Tr} \big[-\beta {\cal P} \, 
\big( {\rm ch}(\beta g A_x) - 1 \big) 
+ \beta g A_x {\rm sh}(\beta g A_x)
- {\rm ch}(\beta g A_x) + 1 \big] \;.
\ee
To further analyze the effective potential (\ref{full-potential}), 
we define the dimensionless potential
$U \equiv \beta (1 + \beta {\cal P}) V_{\rm eff}/\rho$ ,
the dimensionless parameter $c \equiv \beta {\cal P} > 0$ 
and the dimensionless variable $s \equiv \beta g A_x$.
The gauge potential $A_x$ is treated as a scalar,
not matrix, quantity. Then, 
\be
\label{U-potential}
U(s) = (1 + c)(1 - {\rm ch}s ) + s \, {\rm sh}s \;.
\ee
One easily finds
\begin{displaymath}
U(s) \approx \left\{ \begin{array}{ccc} 
\frac{1-c}{2}\, s^2 + \frac{3-c}{24}\, s^4  & {\rm for} & \frac{1}{s} \gg 1 \;,
\\[3mm]
\frac{s-c}{2} \, e^s &  {\rm for} & s \gg 1 \;.\end{array} \right. 
\end{displaymath}
As seen, the potential $U(s)$ always grows for sufficiently large 
$s$. Since the derivative of $U(s)$ equals 
$U'(s) = - c \, {\rm sh}s + s \, {\rm ch}s$, the potential $U(s)$
monotonously grows for $c < 1$ as the equation
$s = c \,{\rm th}s $ has no solution. For $c > 1$ the potential 
has a minimum. For $c$ only slightly exceeding unity {\it i.e.} 
such that $0 < c - 1 \ll 1$, the potential is minimized for 
$s_{\rm min} \approx \sqrt{3 (c -1)}$. For $c \gg 1$ the minimum is 
at $s_{\rm min} \approx c$. The potential (\ref{U-potential}) is 
shown in Fig.~1.
                                                                                

\section{Discussion}


The effective potentials analyzed in the previous sections 
have been derived, assuming that the commutation condition 
(\ref{condition2}) is satisfied. The assumption is discussed 
here in more depth. We also consider the hierarchy of time scales 
of the dynamical evolution of the unstable system. The discussion 
reveals relevance of our results. Let us start with the commutation 
condition (\ref{condition2}).

It has been argued in \cite{Arnold:2004ih} that for the unstable
configuration the color direction of the steepest descent of the 
effective potential corresponds to the Abelian configuration 
where the Yang-Mills term (\ref{YM-potential}) vanishes. Therefore, 
the system becomes Abelian in the course of instability development. 
The suggestion has been confirmed by the numerical simulations 
\cite{Dumitru:2005gp,Arnold:2004ih,Rebhan:2004ur} in $1+1$ dimensions 
which are relevant for our effective potential. (As mentioned in the 
Introduction, the abelianization in $1+3$ dimensions 
\cite{Arnold:2005vb,Rebhan:2005re} is operative only for sufficiently 
small gauge fields.) Thus, there is a good reason to expect that the 
commutation condition (\ref{condition2}) is satisfied. However, in the 
analysis \cite{Arnold:2004ih}, only the Yang-Mills (\ref{YM-potential}) 
and quadratic term have been taken into account. So, one wonders 
whether the argument still holds when the higher order terms are 
taken into account. We note that if the strongly unstable configuration
becomes Abelian at the Hard Loop ($g^2$) level, it remains Abelian
for arbitrary large gauge field amplitudes. There seems to be no
reason to change the runaway direction in the color space which
corresponds to the Abelian configuration. We obviously assume here
that initially the gauge fields are small. Thus, we believe that
the condition (\ref{condition2}) is justified for the strongly unstable 
configurations, as that one corresponding to the Gaussian momentum
distribution. When we deal with the weakly unstable configuration, as that 
one discussed in Sec.~\ref{weak}, the commutation condition (\ref{condition2})
is no longer justified when the effective potential starts to climb up.
Then, the system's color configuration can become non-Abelian again.

Our main objective was to derive the higher order terms of the 
effective potential. Let us give a crude estimate of the time scale 
when the quartic term becomes relevant. For the purpose of order of 
magnitude estimates, we treat the gauge potential as a scalar quantity 
$A$ and we write down the effective potential as
\be 
\label{simple-eff-pot}
V_{\rm eff} \sim - \mu^2 A^2 + \lambda A^4 \;,
\ee
where $\mu^2 > 0$. One immediately observes that when 
$A^2 \sim \mu^2/|\lambda|$, the quartic contribution to the
effective potential is equal to the quadratic contribution.
One estimates $\mu$ and $\lambda$ as
$$
\mu^2 \sim g^2 \frac{\rho}{T} \;, \;\;\;\;\;\;\;
\lambda \sim g^4 \frac{\rho}{T^3} \;,
$$
where $\rho$ is the parton's density and $T$ is, as previously, the 
parton's characteristic momentum or energy. Thus, the quartic term
is important when $A \sim T /g$. At this scale the system's energy, 
which is stored in the fields, is comparable to the total energy of 
particle excitations. When $A \sim T /g$ not only the quartic term 
but the terms of any order are important and the system's dynamics 
becomes nonperturbative.

And how much time is needed for the amplitude of an unstable
mode to become so large that the quartic contribution to the
effective potential is comparable to the quadratic one?
When the instability develops, the gauge potential grows as
$A (t) = a \, e^{\gamma t}$, where $a$ is the initial value
of $A$ and $\gamma$ is the instability growth rate. Substituting
the exponential dependence of $A$ into Eq.~(\ref{simple-eff-pot}),
one finds that the quartic contribution equals the quadratic
one after the time
$$
\label{quartic-time}
t_{\rm higher} \sim \frac{1}{\gamma}
{\rm ln}\Big(\frac{\mu^2}{|\lambda |a^2}\Big) \;.
$$
To get $a^2$, which we treat as a fluctuating gauge potential
in a neutral background, we first estimate the fluctuating current
$j$ which was studied in \cite{Mrowczynski:1996vh}. One finds that
$j^2 \sim g^2 \rho^2 /N$, where $N$ is a number of partons in the
volume which matters for the unstable mode generation. It is identified
with the cubic wavelength of the unstable mode. Thus, $N \sim \rho/k^3$,
where $k$ is the characteristic wave vector of the unstable mode,
and $j^2 \sim g^2 \rho \, k^3$. Physically, the current appears due
to the charge fluctuations in the volume of interest. Further, the
gauge potential squared is estimated as
$a^2 \sim j^2/ k^4 \sim g^2 \rho /k$. Since $k^2 \sim \mu^2$
\cite{Mrowczynski:xv,Randrup:2003cw,Romatschke:2003ms,Arnold:2003rq},
one gets $a^2 \sim g \sqrt{\rho T}$ \cite{Arnold:2004ti}.
Because the instability growth rate $\gamma$ is also of order of $\mu$
\cite{Mrowczynski:xv,Randrup:2003cw,Romatschke:2003ms,Arnold:2003rq},
we finally get
$$
t_{\rm higher} \sim \frac{1}{g} \sqrt{\frac{T}{\rho}}
\: {\rm ln}\Big(\frac{T^3}{g^6\rho}\Big) \;.
$$
If $\rho \sim T^3$, as in the equilibrium, the estimate
simplifies to
$$
t_{\rm higher} \sim \frac{1}{g T}
\: {\rm ln}(1/g) \;.
$$

Let us now consider the characteristic time of back reaction 
($t_{\rm back}$) which is defined as a time interval after which 
the change of parton's momentum ($\Delta p$) due to the Lorentz 
force ($F$) is comparable to the parton's momentum itself ($T$). 
One computes 
$$
\Delta p \sim \int_0^{t_{\rm back}} F(t)dt 
\sim gk \int_0^{t_{\rm back}} A(t)dt
\sim \frac{gka}{\gamma} \,\big( e^{\gamma t_{\rm back}} - 1\big) \;.
$$
Neglecting the unity in the last expression and using the above 
estimates of $k$, $a$ and $\gamma$, the equation $\Delta p = T$ 
provides
\be
\label{back-time}
t_{\rm back} \sim \frac{1}{g} \sqrt{\frac{T}{\rho}}
\: {\rm ln}\Big(\frac{T^3}{g^6\rho}\Big) \sim t_{\rm higher} \;.
\ee
As seen, the time scale of back reaction and that of relevance of 
higher order terms of the effective potential are the same. This is 
not an accident - the two effects are actually the same. We 
could define the scale of back reaction time as the time when the 
parton distribution function which, according to Eqs.~(\ref{ansatz}), 
is a function of $p \pm gA$, significantly depends on $gA$. And then,
it is evident that $t_{\rm back} = t_{\rm higher}$.

The shortest time scale of interest is that of instability growth
and consequently, the early stage of instability development can
be studied with the frozen momentum distribution. At the time scale 
of back reaction the influence of the mean field on the parton's motion 
cannot be longer neglected. At this scale the anisotropic system is 
expected to evolve fast to isotropy. The time scale of inter-parton 
collisions controls the local equilibration of the system. However, 
the scales of both hard and soft collisions, given by, respectively, 
Eq.~(\ref{hard-time}) and Eq.~(\ref{soft-time}), are much longer 
than $t_{\rm back}$ (\ref{back-time}). A plasma system produced in 
relativistic nucleus-nucleus collisions is  subject to expansion due 
to the system's finite size, an effect that we have not considered. 
It should be noted that the expansion provides another relevant time 
scale in the problem. As shown in \cite{Randrup:2003cw}, it is so 
fast that the momentum distribution evolution caused by the expansion 
can stop the instability growth. 

We conclude our considerations as follows. Configurations of an 
anisotropic quark-gluon plasma, which are unstable within the linear 
response approach, can be strongly or weakly unstable, depending on 
the sign of higher order terms of the effective potential. The sign 
in turn depends on the parton momentum distribution. Examples of both 
situations have been given. As argued above, the non-Abelian non-linearities 
do not stop the instability development if the momentum configuration is 
strongly unstable. To quantitatively understand the system's dynamics 
at longer times, one has to go beyond the Hard Loop approximation, 
which effectively amounts to consider the effect of back reaction of 
the mean fields on the parton's motion. At even longer times, the 
effect of inter-parton collisions, which drive the system towards 
local thermal equilibrium, has to be taken into account. Studies of
finite systems require a proper treatment of the expansion which may
affect seriously the instability development.

\acknowledgements

We are grateful to Peter Arnold for a correspondence on the fluctuating 
fields, and to Mikko Laine for discussions. C.~M. thanks IEEC for 
hospitality during completion of this work. Financial support by MEC 
(Spain) under grant FPA2004-00996 is also acknowledged.

\appendix


\section{}
\label{appendix1}


Let us discuss here the quadratic term of the  lagrangian 
(\ref{efflag}). The term equals \be
\label{efflag-(2)}
{\cal L}_{\rm eff}^{(2)}(x) = - \frac{1}{2} 
{\cal K}^{\mu \nu} A_{\mu}^a (x) A_{\nu}^a (x) \;,
\ee
where
\be
\label{K} 
{\cal K}^{\mu \nu} \equiv 
\frac{g^2}{2} \int dP \: p^{\mu} \:
\frac{\partial  f(p)}{\partial p_{\nu}} \;.
\ee
Only the quark contribution is taken into account here. 
The term (\ref{efflag-(2)}) should be compared to the lagrangian 
derived in the linear response analysis which is
\be
\label{LR-action}
{\cal L}_{\rm LR}(x) =  - \frac{1}{2} 
\int d^4y A^a_\mu(x) \Pi^{\mu \nu}(x-y) A^a_\nu(y) \;, 
\ee
where the Fourier transformed polarization tensor $\Pi^{\mu \nu}(k)$ 
of anisotropic plasma equals \cite{Mrowczynski:2000ed}
\be
\label{Pi}
\Pi^{\mu \nu}(k) = \frac{g^2}{2} \int dP \: p^{\mu}
{\partial f (p) \over \partial p_{\lambda}} \Bigg[ g^{\lambda \nu} 
- { k^{\lambda} p^{\nu} \over p^{\sigma}k_{\sigma} + i0^+} \Bigg] \;.
\ee
As in the case of ${\cal K}^{\mu \nu}$, only the quark contribution 
has been written down. We note that $\Pi^{\mu \nu}(k) = \Pi^{\nu \mu}(k)$ 
and $k_\mu \Pi^{\mu \nu}(k) =0$. We also observe that the Hard Loop 
action was found for both the equilibrium 
\cite{Braaten:1991gm,Frenkel:1991ts,Kelly:1994dh,Blaizot:1993be} 
and anisotropic \cite{Mrowczynski:2004kv} plasmas, modifying 
the action (\ref{LR-action}) to comply with a gauge invariance. 

One finds that the coefficient (\ref{K}) can be written down as
$$
{\cal K}^{\mu \nu} = \Pi^{\mu \nu} (\omega, {\bf k}=0) \;,
$$ 
where $k \equiv (\omega, {\bf k})$. It is worth noting that 
$$
\lim_{\omega \rightarrow 0} \Pi^{\mu \nu} (\omega, {\bf k} = 0) 
\not=
\lim_{{\bf k} \rightarrow 0} \Pi^{\mu \nu} (\omega = 0, {\bf k}) \;.
$$ 
We also observe that $\Pi^{\mu \nu} (\omega, {\bf k}=0)$ is 
independent of $\omega$.

There is a subtle point concerning the calculations of 
$\Pi^{\mu \nu}$ and of ${\cal K}^{\mu \nu}$. Because of its 
transversality, the polarization tensor (\ref{Pi}) can be 
computed in two different but equivalent ways: either the 
distribution function $f(p)$ is treated as a function of all 
four components of the four-momentum $p$ or we substitute into 
Eq.~(\ref{Pi}) the function $f({\bf p})= f(E_p,{\bf p})$ which 
depends only on the three-momentum. When ${\cal K}^{\mu \nu}$ 
is computed, the two methods are not equivalent to each other. 
However, the derivation of the effective potential (\ref{efflag}) 
shows that $f(p)$ has to be treated here as a function of all four 
independent components of $p$.  

\section{}
\label{appendix2}

We present here some details of the calculations which lead to the 
effective potential (\ref{stabil-potential}) of weakly unstable
system.

The normalization constant $C$ of the distribution 
(\ref{stabil-distri}) is determined by the integral 
$$
\rho = \int \frac{d^4p}{(2 \pi)^3} 2\Theta(p_0)\, \delta(p^2) \,
p_0 \, f(p_x, p_y, p_0) = C  \int \frac{d^3p}{(2 \pi)^3} 
\big[ \delta (p_x - {\cal P}) + \delta (p_x + {\cal P}) \big]
\, e^{- \beta E_p} \;,
$$  
where $E_p \equiv \sqrt{p_x^2 + p_y^2 + p_z^2}$. It can be easily
computed, using the variables $(p_x,E,\phi)$ with $\phi$ being 
the azimuthal angle. Then, one finds
$$
\rho = \frac{C}{(2 \pi)^2} \int_{-\infty}^{\infty}dp_x
\big[ \delta (p_x - {\cal P}) + \delta (p_x + {\cal P}) \big]
\int_{|p_x|}^{\infty}dE_p \,E_p \, e^{- \beta E_p}
= \frac{C}{2 \pi^2 \beta^2} \, 
e^{-\beta {\cal P}}(1 + \beta  {\cal P}) \;.
$$

To get the effective potential (\ref{stabil-potential}), we first compute
$$
\Big\langle \frac{p_x}{p_0} \frac{\partial}{\partial p_x}
\Big\rangle 
= \int \frac{d^4p}{(2 \pi)^3} 2\Theta(p_0)\, \delta(p^2) \,
p_0 \; \frac{p_x}{p_0} \frac{\partial}{\partial p_x}
f(p_x, p_y, p_0) 
=\int \frac{d^3p}{(2 \pi)^3} \frac{p_x}{E_p} 
\frac{\partial}{\partial p_x} f(p_x, p_y, E_p) \;.
$$
It should be stressed that the derivative with respect to $p_x$ acts only 
on the explicit dependence of $f(p_x, p_y, E_p)$ on $p_x$. Therefore, 
after the partial integration the derivative acts on $p_x /E_p$  
and on the $E_p$ dependence of $f(p_x, p_y, E_p)$. Then, one finds
$$
\Big\langle \frac{p_x}{p_0} \frac{\partial}{\partial p_x}
\Big\rangle 
= - \frac{\rho \, \beta \, e^{\beta {\cal P}}}{1 + \beta {\cal P}}
\int_0^{\infty}dp_x \,
\delta(p_x - {\cal P}) \,
\frac{\partial }{\partial p_x} \Big( p_x \, e^{- \beta p_x} \Big) 
= - \rho \beta \, \frac{1 - \beta {\cal P}}{1 + \beta {\cal P}}\;.
$$

It is not difficult to generalize the above result 
to the higher order terms. Keeping in mind that
$$ 
\frac{\partial^l }{\partial p_x^l} \Big( p_x \, e^{- \beta p_x} \Big) 
= (-\beta )^{l-1} e^{- \beta p_x} (l - \beta p_x ) \;,
$$
one finds for $l=1,3,5 \dots$ the following result
$$
\Big\langle \frac{p_x}{p_0} \frac{\partial^l}{\partial p_x^l}
\Big\rangle 
= - \rho \,\beta^l 
\frac{l - \beta {\cal P}}{1 + \beta {\cal P}}\;,
$$
which gives the effective potential (\ref{stabil-potential}).

%
%

\end{document}